\documentclass[12pt]{article}

\usepackage{lipsum} 
\usepackage[round]{natbib}
\usepackage{graphicx}
\usepackage[all]{xy}
\RequirePackage{lineno}
\usepackage[toc,page]{appendix}
\usepackage{setspace}

\usepackage[sc]{mathpazo} 
\usepackage[T1]{fontenc} 
\usepackage{microtype} 
\usepackage{colortbl}
\usepackage{multicol} 
\usepackage[hang, small,labelfont=bf,up,textfont=it,up]{caption} 
\usepackage[hmarginratio=1:1,top=15mm,columnsep=20pt]{geometry} 
\usepackage{booktabs} 
\usepackage{float} 
\usepackage{hyperref} 
\usepackage[caption = false]{subfig}
\usepackage{lettrine} 
\usepackage{paralist} 
\usepackage{longtable}
\usepackage[dvipsnames]{xcolor}
\usepackage{abstract} 
\usepackage{pdflscape}
\usepackage{lscape}
\usepackage{amsmath}
\usepackage{array}
\usepackage{tabularx}
\usepackage{array}

\begin{document}

\textbf{ Species traits and community properties explain species extinction effects on detritus-based food webs} \newline

\vspace{0.5cm}

\textbf{Authors:} Idaline Laigle$^{1,2}$, Isabelle Aubin$^{3}$, Dominique Gravel$^{1,2}$ \\
\textbf{1}: D\'{e}partement de biologie, Universit\'{e} de Sherbrooke, 2500 Boulevard l'Universit\'{e},
Sherbrooke (Qu\'{e}bec, Canada). J1K 2R1\\
\textbf{2}: Qu\'{e}bec Centre for Biodiversity Science\\
\textbf{3}: Natural Resources Canada, Canadian Forest Service, Great Lakes Forestry Centre, 1219 Queen St. East, Sault Ste. Marie, ON P6A 2E5, Canada \\

\vspace{0.5cm}
 
\textbf{E-mail addresses of all authors :} \\
idaline.laigle@usherbrooke.ca - ORCID: 0000-0003-1422-9984 \\
isabelle.aubin@canada.ca \\
dominique.gravel@usherbrooke.ca - ORCID: 0000-0002-4498-7076 \\

\vspace{0.5cm}

\textbf{Name and complete mailing address :} Idaline Laigle \\
819-821-8000 -- 62638 \\
D\'{e}partement de biologie, Facult\'{e} des Sciences, Universit\'{e} de Sherbrooke,
2500 Boulevard Universit\'{e}, Sherbrooke, Quebec, Canada J1K 2R1. \\

\newpage
\doublespacing


\begin{abstract}

Effects of changes in functional composition of soil communities on nutrient cycling are still not well understood. Models simulating community dynamics overcome the technical challenges of conducting species removal experiments in the field. However, to date, available soil food web models do not adequately represent the organic matter processing chain which is key for soil dynamics. Here, we present a new model of soil food web dynamics accounting for allometric scaling of metabolic rate, ontogeny of organic matter, and explicit representation of nitrogen and carbon flows. We use this model to investigate what traits are best predictors of species effects on community productivity and on nutrient cycling. To do so, we removed 161 tropho-species (groups of functionally identical species) one at a time from 48 forest soil food webs, and simulated their dynamics until equilibrium. We assessed tropho-species removal effects as the relative changes between the biomass of each component (consumers, detritus, producers, microbes and nitrogen) before, and after the removal. Simulations revealed that combinations of traits better determine removal effects than single ones. The smallest species are the most competitive ones, but carnivores of various body masses presenting the highest connectivity and resource similarity could be key stone species in the regulation of competitive forces. Despite this, most removals had low effects, suggesting functional redundancy provides a high resistance of soil food webs to single tropho-species extinction. We also highlight for the first time that food web structure and soil fertility can drastically change species effects in an unpredictable way. Moreover, the exclusion of detritus and stoichiometric constraints in past studies lead to underestimations of indirect effects and retroactions. While additional work is needed to incorporate complementarity between detritivores, it is essential to take into account these mechanisms in models in order to improve the understanding of soil food web functioning.

\end{abstract}

\section*{Introduction}

Soil communities are largely under-documented despite their importance in the provision of several ecosystem services related, in particular, to soil fertility \citep{Bardgett2010}. Soil food webs are based on detritus and have a distinct functioning from other types of food webs \citep{Moore2004a, Digel2014}. Indeed, detritus production is donor-controlled whereas living prey reproduce and may have several behavioral responses to predation. This characteristic and the wide diversity of life forms within soil community lead to important feedbacks, and make difficult the study of relationships between soil food web structure and functioning. Theoretical studies and mesocosm experiments investigated the role of several taxonomic groups in the process of litter decomposition, but so far, we do not know which species properties best predict the way they affect this process. Functional traits can reveal generalities across ecosystems \citep{McGill2006, Violle2007a,Cardinale2012}, and provide a common currency to reduce the tremendous diversity of soil food webs. It is therefore legitimate to ask if the trait based approach could be used to better understand how changes in community structure could alter ecosystem functions \citep{Wardle2006,Bardgett2010}.

A species effect on ecosystem functions, such as productivity or nutrient cycling, can be direct, caused by a particular property involved in a specific function. For example, a grazer can efficiently regulate bacteria biomass, and the vertical movement of earthworms distributes organic matter in the soil profile. Species effects on ecosystem functions can also be indirect via trophic regulation that propagates throughout the network of interactions \citep{Montoya2009}. A species effect on other species can cascade from the bottom to the top of the food chain, such as the plant composition of a community that can affect the abundance and the composition of herbivores, detritivores, predators and parasitoids \citep{Hawes2003, Bohan2005a}. Alternatively, looking from the top to the bottom, \citet{Schmitz2003,Schmitz2009} showed how predator functional type, mostly determined by hunting strategy, influences herbivores behavior, plant composition, and consequently nutrient cycling. Indirect effects can also be observed between species due to apparent or exploitative competition for a shared consumer or resource, respectively. For instance, \citet{Montoya2009} showed that predators increased the equilibrium biomass of their prey within soil web models instead of decreasing it for 40\% of predator-prey interactions. Indirect effects in food webs, although being often counter-intuitive, are common and can significantly influence population dynamics. Consequently, several authors \citep{Raffaelli2002, Werner2003, Tylianakis2008, Brose2016} have suggested that a multi-trophic assessment of community dynamics is required to understand processes involved in community assembly and ecosystem functioning.

Models of community dynamics are powerful tools to investigate theoretically how species can directly and indirectly affect each other, and how they could affect the ecosystem functioning \citep{Duffy2007, Berg2015, Gravel2016}. Numerous models of soil community dynamics have been built. For instance, \citet{Hunt1987, Hunt2002} and \citet{DeRuiter1993} used models to explore the role of different taxonomic groups in the mineralization process. These models provided some insight on soil functioning, but they lacked precision and reproducibility \citep{Scheu2002,Buchkowski2016}. These models were based on food webs composed of coarse taxonomic groups, using linear feeding relationships and parameters that are difficult to extrapolate to different species and food webs. Alternatively, \citet{Yodzis1992} and \citet{Brose2006} developed the Allometric Trophic Network (ATN) model, which is based on allometric relationships \citep{Peters1983,Brown2004} to represent energy flows within food webs. The ATN model gains in generality building on the universal scaling of metabolic rates with body mass, but lacks realism for soils by neglecting key aspects of soil community dynamics such as stoichiometry \citep{Daufresne2001}, the ontogeny of organic matter decomposition \citep{Moore2004a} and slow-fast channels \citep{Rooney2006}. Stoichiometric constraints improve the representation of consumer-resource dynamics \citep{Elser2000}, and provide a more rigorous assessment of biochemical processes by accounting for nutrient imbalances \citep{Woodward2002, Elser1999}. The progressive modification of organic matter C:N is also a key feature of the organic matter processing chain in soil food webs but is rarely included in models of community dynamics \citep{Moore2004a}. A model incorporating those aspects of soil dynamics may improve the investigation of species contribution to nutrient cycling, and may provide some insight on general rules determining species extinction effects in soil food webs.

While it is obvious that the abiotic environment has a significant effect on species demography, environmental conditions are rarely taken into account in models investigating species contribution to ecosystem functioning. Soil fertility, related to litter quality, which can be assessed with nitrogen availability, has important effects on soil community and ecosystem processes \citep{Wardle2004}. For instance, the relative importance of microbes in organic matter decomposition changes according to litter properties \citep{Wardle2004,DeGraaf2010}, fungi being more effective than bacteria in the decomposition of recalcitrant litter than labile one, and inversely. Soil organisms also have different effects according to community composition and soil fertility \citep{Scheu2002}. For example, a fungivore should have different effects according to the proportions of fungi and bacteria, and therefore soil fertility. In addition, if prey of low C:N are present in a community, their consumers should have positive effects on soil fertility by excreting nitrogen. Investigating different food web structure, species stoichiometry and soil fertility is then a good way to assess whether species effects vary according to environmental conditions \citep{Buchkowski2016}.

To our knowledge, no study investigated how species functional traits explain species effects on soil ecosystem functions in a multi-trophic context. The objective of our study is to assess whether we could explain species effects on the food web from its traits and trophic position. We developed a model of soil community dynamics in which carbon and nitrogen flows are determined by stoichiometric constraints, and metabolic rates are calculated from species body mass. We considered the topology of 48 forest soil food webs previously documented in Germany along with the functional traits of 878 species, that comprise them, gathered in tropho-species. We performed numerical experiments in which we removed each tropho-species one at a time, and then ran the model until equilibrium. We calculated the effects of these removals on the biomass of consumers, producers, microbes, detritus, and on inorganic nitrogen. \textit{Per population}, and \textit{per capita} effects of tropho-species were modeled as a function of their functional and trophic properties. We repeated the simulations with a different level of available inorganic nitrogen to assess whether species effects vary with soil fertility. Since body mass and C:N ratio affect consumption and excretion rates, we expected these to be important variables explaining species effects on nitrogen mineralization and overall community productivity. Following the niche partitioning theory \citep{Macarthur1958,Hardin1960}, species functionally close to each other tend to compete more than functionally distinct species. We therefore hypothesized that species that are functional redundant to the other ones will be detrimental to consumers biomass. In addition, functionally unique species should have greater effects on the overall food web than redundant ones. And finally, according to \citet{Sole2001,Dunne2002} and \citet{Eklof2006}, we expected that the loss of the most connected species, known to stabilize food webs, will have higher effects on the overall community than less connected ones.

\section*{Methods}

We ran numerical experiments using an empirically-derived model that simulates community dynamics of diverse soil food webs with changing species composition. 

\subsection*{Dynamics}

\begin{figure}[H]
\begin{center}
\includegraphics[width=0.6\textwidth]{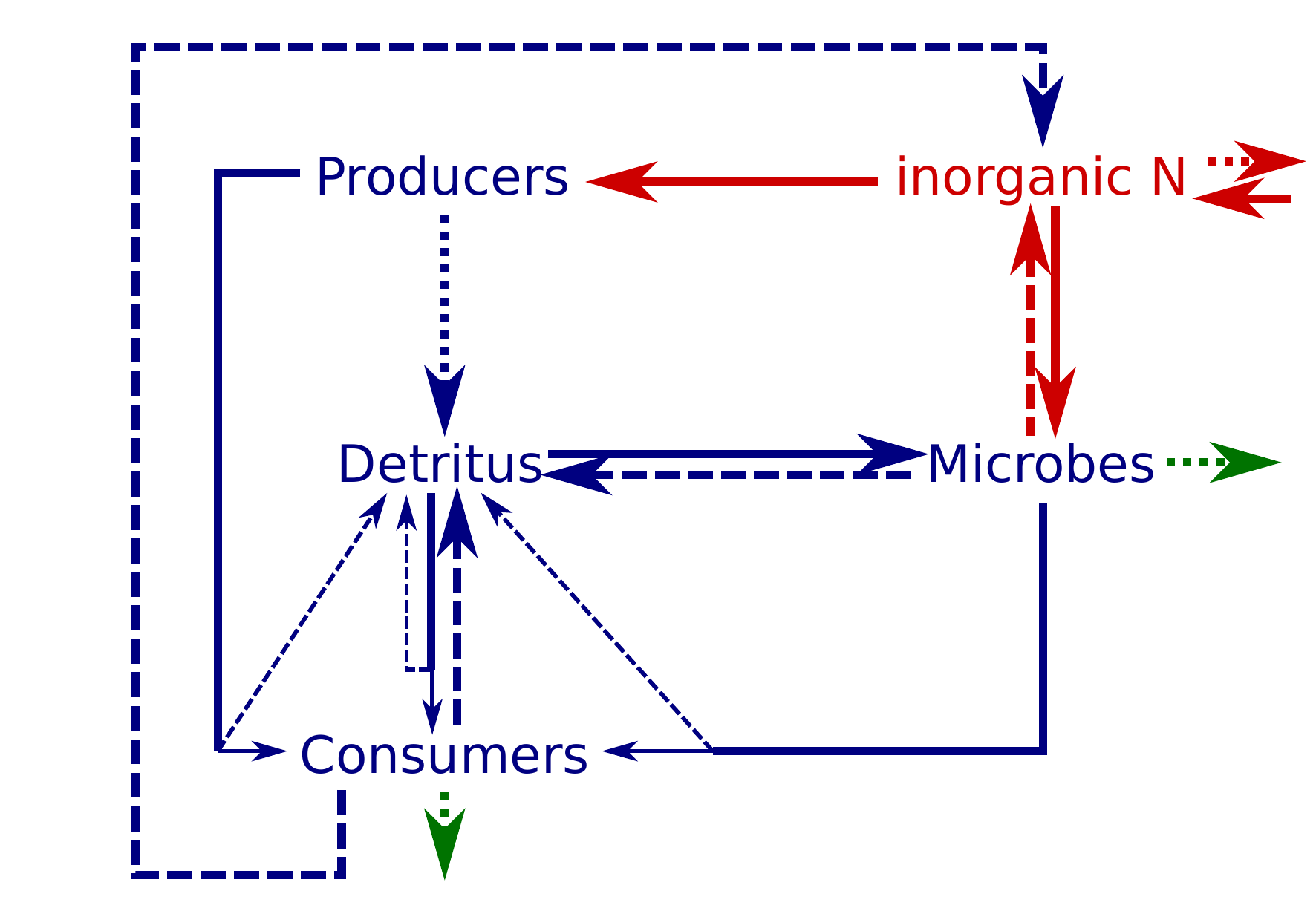}
\caption{Diagram of flows and components considered in the model. Carbon flows are represented in green, nitrogen in red, and both in blue. Dotted arrows represent litter loss, respiration and leaching; dashed arrows represent excretion and plain arrows represent consumption or absorption.}
\label{plan}
\end{center}
\end{figure}

The model consists of five compartments each with their different dynamics : producers (P), detritus (D), inorganic nitrogen (N), microbes (bacteria and fungi - M) and consumers (E). Biomass flows are divided between carbon and nitrogen (Figure \ref{plan}). We present below the equations used to describe carbon biomass variation over time, and inorganic nitrogen dynamic. All variables are expressed in mass per unit area. Estimated parameters are presented in table \ref{param} with relevant references.

\textbf{Producers} assimilate inorganic nitrogen and carbon dioxide, and lose biomass by leaf senescence and consumption. We considered that producers were not limited by light, water or $CO^2$. Producer growth is described using a Monod equation \citep{Monod1949} following \citet{Tilman1982} and \citet{Daufresne2005}, and is thus nitrogen limited. The model represents both the biomass of underground and aerial parts. Producer biomass production and loss are thus described as: \\

\begin{equation}
\frac{dP_p}{dt} = r_p \cdot P_p \cdot \frac{N}{K_p + N} - (l_p \cdot P_p) - \sum_{i=consumers}S_{pi},
\end{equation}

where $r_p$ is the asymptotic growth rate of producer species $p$, $P_p$ is producer $p$ biomass in units of carbon, $N$ is the inorganic nitrogen available in the soil, $K_{p}$ is the half saturation constant for nitrogen of producer $p$, $l_p$ is the litter loss rate of producer $p$ and $S_{pi}$ is the biomass of $p$ consumed by consumer species $i$. The amount of nitrogen absorbed by producer $p$ ($A_{Np}$) is proportional to carbon assimilated to maintain producer C:N. Litter produced has the same C:N than the producer. \\

\textbf{Microbes} consume only detritus. Growth of the different microbial species $m$ follows a modified Monod equation \citep{Holmberg1982, Blagodatsky1998, Moorhead2006}, and they lose carbon by respiration and predation: \\

\begin{equation}
\frac{dM_m}{dt} = r_m \cdot M_m \cdot \sum_{d=detritus} \frac{f_{md} \cdot D_d}{K_{m} + \sum\limits_{d=detritus} f_{md} \cdot Dd} - (g_m \cdot M_m) - \sum_{i=consumers}S_{mi},
\end{equation}

where $r_m$ is microbe $m$ asymptotic growth rate, $M_m$ is microbe $m$ biomass in units of carbon, $f_{md}$ is microbe $m$ preference for detritus $d$, $D_d$ is the biomass of detritus $d$, $K_m$ is microbe $m$ half saturation constant for detritus, $g_m$ is microbe $m$ respiration rate, and $S_{mi}$ is the biomass of microbe species $m$ consumed by consumer species $i$. Preference for detritus decreased linearly with increasing detritus C:N as C:N is correlated to palatability \citep{Heal1997}. The sum of the preferences of one microbe equals to one. \\

\textbf{Consumers} feed on detritus, producers, microbes or on other consumers. Consumption rates are calculated following the work of \citet{Yodzis1992} and \citet{Brose2006}. As suggested by \citet{DeRuiter1993}, consumers convert into their own biomass only a part of the biomass they consume according to their assimilation efficiency ($e_{j}$), which depends on type of food consumed (animal or non-animal), while the rest is excreted and respired. Consumers dynamics are represented by: \\

\begin{equation}
\frac{dE_i}{dt}= E_i \cdot y_i \cdot z_i \cdot \sum_{j=prey} e_j \cdot (\frac{f_{ij} \cdot R_j^h}{B_0^h+\sum\limits_{w=prey} f_{iw} \cdot R_w}) - (E_i \cdot z_i) - \sum_{j=consumers}S_{ij},
\end{equation}

where $E_i$ is biomass of consumer $i$, $y_i$ is the maximum consumption rate relative to metabolic rate of $i$, $R_j$ is the biomass of resource $j$, $B_0$ is the half-saturation density, $h$ is the Hill exponent equals to 2 to obtain a type III functional response, $S_{ij}$ is the biomass of $i$ consumed by consumer species $j$, $f_{ij}$ is species $i$ preference for species $j$. $f_{ij}$ is given by the inverse of the number of prey of consumer $i$. When consumer $i$ is a herbivore or a detritivore, their preferences for detritus or plants decreased linearly with increasing detritus and plants C:N as C:N is correlated to palatability \citep{Heal1997}. The sum of the preferences of one consumer equals to one.

$z_i$ is the mass specific metabolic rate of consumer $i$ calculated as: \\
 
\begin{equation}
z_i = x_i \cdot {Q_i}^{c},
\end{equation}

where $Q_i$ is the body mass of consumer $i$, $x_i$ is the allometric constant of consumer $i$, $c$ is an allometric exponent assessed according to \citet{Reuman2009}. \\

\textbf{Detritus} inputs come from leaf loss and from microbes and consumers excretions, while leaching and consumption by microbes and consumers are losses: \\

\begin{equation}
\frac{dD_d}{dt} = \sum_{i=consumers}H_i + \sum_{p=producers}(l_p \cdot P_p) + \sum_{m=microbes}H_m - \sum_{i=consumers}S_{di} - \sum_{m=microbes}S_{dm} - b_d \cdot D_d,
\end{equation}

where $D_d$ is the biomass of detritus $d$, $H_i$ and $H_m$ are the inputs from consumers $i$ and microbes $m$, respectively, $S_{di}$ is the consumption of detritus $d$ by consumers $i$, $S_{dm}$ is the consumption of detritus $d$ by microbes $m$, and $b_d$ is the leaching rate of detritus $d$. Detritus are divided into five different pools according to their quality \citep{Moore2004a}. Detritus quality is defined by C:N ratio: $1<C:N<10$, $10<C:N<20$, $20<C:N<35$, $35<C:N<50$, No N (humus). The C:N ratio of the input is calculated to know which pool will be filled. At the end of each time step the new C:N ratio of detritus is updated to keep mass balance. We suppose a constant leaf input with a C:N ratio of 26, corresponding to the mean C:N ratio of detritus in temperate forest litter \citep{Gloaguen1982}. \\

\textbf{Consumers and microbes stoichiometric regulation:}

The model considers stoichiometry of C and N explicitly assuming a fixed homeostasis in consumers and microbes C:N, as in \citet{Daufresne2001}. Microbes and consumers excrete nitrogen or carbon when in excess in the consumed resources \citep{Mclaren1996}, but if inorganic nitrogen is available when carbon is in excess, microbes absorb nitrogen. For microbes we consider carbon excretion as humus (pure carbon) while excreted nitrogen fills the inorganic nitrogen pool. For consumers, surplus of carbon or nitrogen are added to the non assimilated part of the consumed prey biomass ($H_i$), and excreted into the corresponding detritus pool.  \\

Excreted nitrogen by consumers and microbes is calculated as follow:  \\

\begin{equation}
X_{Ni/m} = O_{Ni/m} - \frac{O_{Ci/m}}{C:N_{i/m}},
\end{equation}

where $O_{Ni/m}$ is total consumed nitrogen by consumer $i$ or microbe $m$, $O_{Ci/m}$ is total consumed carbon minus metabolic losses, and $C:N_{m/i}$ is the C:N ratio. \\

Excreted carbon by consumers $i$ and microbes $m$ is calculated as follow: \\

\begin{equation}
X_{Ci/m} = O_{Ci/m} - O_{Ni/m} \cdot C:N_{i/m},
\end{equation}

Absorbed nitrogen by microbes $m$ is calculated as follow : \\

\begin{equation}
A_{Nm} = \frac{O_{Cm}}{C:N_m} - O_{Nm},
\end{equation}

\textbf{Inorganic nitrogen} is lost by leaching, producer and microbial absorption, and is replenished by atmospheric deposition, as well as excretion from consumers and microbes: \\

\begin{equation}
\frac{dN}{dt} = \sum_{i=consumers}X_{Ni} + \sum_{m=microbes}X_{Nm} - \sum_{p=producers}A_{Np} - \sum_{m=microbes}A_{Nm} + T - h \cdot N,
\end{equation}

where $X_{Ni}$ is nitrogen excreted by consumer $i$, $X_{Nm}$ is nitrogen excreted by microbe $m$, $A_{Np}$ is nitrogen absorbed by producer $p$, $A_{Nm}$ is nitrogen absorbed by microbe $m$, $T$ is the amount of atmospheric deposition and $h$ is the leaching rate of nitrogen. \\

The model is written in C++ and the numerical integration performed with the Runge Kutta Fehlberg 78 method \citep{Fehlberg1969}, using the library 'odeint' \citep{odeint}.

\singlespacing

\begin{table}[H]
\centering
\begin{tabular}{p{1.8cm} p{1.6cm} p{1.5cm} p{2.0cm} p{4.5cm} p{3.4cm}}
\hline
Component & Parameter & Unit & Value & Details & Source \\ \hline
Producers & $r_p$ & $time^{-1}$ & 5,50,64,96 & Producers with the lowest C:N are more efficient in nitrogen rich soil, and vice versa. & \\
 & $K_p$ & $mass$ & 1,2,10,30 &  & \\
 & $l_p$ & $time^{-1}$ & 0,0.5,0.4,0.2 & Producers with the higher growth rate lose more leaves & \\
Microbes (fungi, bacteria) & $r_m$ & $time^{-1}$ & 668,512,256, 1536,1024,512 & Microbes with the lowest C:N are more performant when detritus are rich in N, and vice versa.& \\
& $K_m$ & $mass$ & 300,125,25, 125,300,25 &  & \\
& $g_m$ & $time^{-1}$ & 0.01,0.01,0.01, 0.1,0.1,0.1 & Fungi have lower turn rates over than bacteria & \citep{Strickland2010} \\
Consumers & $y_i$ & $time^{-1}$ & 8 & Determined for invertebrates & \citep{Brose2006} \\
 & $B_0$ & $mass$ & 0.5 & Uniform relative consumption rate & \citep{Brose2006} \\
 & $h$ & & 2 & Functional response III & \citep{Real1977} \\
 & $e_i$ & & 0.85,0.65 & For animal resource and others, respectively & \citep{Brose2006} \\
 & $x_i$ & & 0.314 & Determined for invertebrates & \citep{Brose2006} \\
 & $c$ & & -0.25, 0.06, 0.96 & For bodymass $>1.10^{-5}$, $<1.10^{-5}$ and $>1.10^{-7}$, $<1.10^{-7}$, respectively & \citep{DeLong2015} \\
Detritus & $b_d$ & $time^{-1}$ & 0.01 & Random & \\
Nitrogen & $T$ & $mass$ & 0.1 & Random & \\
 & $h$ & $time^{-1}$ & 0.1 or 0.5 & Random & \\
\hline
\end{tabular}
\caption{Estimated parameters used in the model.}
\label{param}
\end{table}

\doublespacing

\subsection*{Food web data}

We used 48 forest soil food webs documented by \citet{Digel2014} to define the network structure, along with species body mass to parameterize the model. These food webs were inventoried in beech and coniferous forests in Germany. Interactions were detected using a combination of methods ranging from molecular gut content analyses to cafeteria experiments (details in \citet{Digel2014}). The smallest species and those at low trophic levels were not identified to the species levels (e.g. nematodes). We divided detritus, producers and microbes into five, four and six groups, respectively. Microbes were further divided into three bacteria and three fungi. The various groups have variable C:N ratios. Growth rates of producers and microbes increase inversely with their C:N ratio \citep{Garnier2004,Keiblinger2010}. Producers and microbes with a low C:N ratio are more efficient in N rich systems, and species with high C:N are more efficient in N poor systems. C:N ratio of fungi and bacteria were documented from \citet{Mouginot2014}, and C:N ratio of producers were documented from the TOPIC database (Traits of Plants in Canada, \citep{Aubin2012}). Metabolic rates of consumers were estimated using their body mass originally measured by \citet{Digel2014}, following the negative-quarter power law relationship with body mass \citep{Peters1983,Delong2010}. The C:N ratios of consumers were assessed according to \citet{Hunt1987} and \citet{Crotty2014}. 

We simulated community dynamics of the food webs with various tropho-species removals to test their effects on the food web. Tropho-species were composed of one to 94 functionally identical consumer species, for a total of 161 tropho-species. Each food web was then comprised of 46 to 71 consumer tropho-species. Traits considered were body mass, soil vertical position, mobility, toughness, use of poison to hunt, use of web, and diet (table \ref{traits} - from \citet{Laigle2018}). We added species order (taxonomic rank) as a proxy of latent traits that are not measurable or which are the result of several traits (i.e. behavior, chemical defenses) \citep{Rohr2010, Mouquet2012, Laigle2018}. Because of the high inter-specific variability in body mass, species were categorized into 10 body mass classes. Tropho-species were then comprised of species belonging to the same order, body mass class, and having identical trait values. We considered that pairs of tropho-species interacting with each other in at least one food web also interacted with each other in all food webs where they co-occurred.

\subsection*{Simulations and analyses}

We simulated community dynamics of the 48 food webs until equilibrium (reached before 150 time steps), then we removed tropho-species one at a time and ran again the simulation until equilibrium (50 extra time steps). We then calculated tropho-species removal effect on other components (fungi, bacteria, detritus, producers, consumers and nitrogen). The effect of a tropho-species $j$ on component $i$, was considered as the inverse of its removal effect $A_{ij}$, calculated as the relative change in mass of each component induced by the removal. This effect was calculated as :

\begin{equation}
A_{ij} = \frac{BR_{ij}- B0_i}{B0_i},
\end{equation}

where $B0_i$ is the biomass of component $i$ at equilibrium in the entire community, and $BR_{ij}$ is the biomass of component $i$ at equilibrium in the community without species $j$. Positive value of $A_{ij}$ indicates that the removed species $j$ has a negative effect on the component $i$, while a negative value indicates a positive effect. We also calculated \textit{per capita} effects by dividing $A_{ij}$ by the biomass of the removed species ($B0_j$) in the full community. 

We investigated how $A_{ij}$ and $A_{ij}/B0_j$ are determined by trophic and functional properties of the removed species using Random Forest algorithms (RF) \citep{Breiman2001}. RF is a machine learning algorithm, which, using decision tree-like processes, finds the best combination of variables and variable coefficients to explain the response variable. We used this method because it effectively handles non-linear relationships and combinations of explanatory variables we wanted to describe. We assessed the importance of each species properties in the explanation of their \textit{per population} and \textit{per capita} effects on the five components. This investigation was inspired by the results of \citet{Laigle2018} who have shown that species traits affect species interactions and thus, should affect species effects in food webs. We also looked at how species trophic position could  determine species effects through their interactions. Species properties that were used as explanatory variables were therefore: removed species traits (table \ref{traits}), mean and minimum functional distance, mean trophic similarity, number of resource and number of consumer. We estimated functional distance between species by calculating Gower distances \citep{Gower1966} on the matrix of species traits and taxonomy. Taxonomy was considered as the scores of each species from the two first axes of the Principal Coordinates Analysis (PCoA) conducted on taxonomic distances between species obtained with the R package "ade4" \citep{Dray2007}. We added an index of taxonomic distance to overcome the lack of time calibrated phylogenies and the challenge of accounting for a wide range of organisms simultaneously (from bacteria to arthropods). To calculate functional distances, the sum of the relative weight of each variable equaled one when they were divided into more than one column (e.g. diet). Trophic similarity was calculated as the percentage of shared resources and consumers for each pair of species. We also used food web identity as an explanatory variable to detect whether species effects were dependent on the food web in which they were. To assess RF accuracy, we calculated the adjusted $R^2$ between observed effects and predicted effects (Rpo). Additionally, we investigated pair relationships between significant species property and tropho-species effects on each component to better understand RF results. With this work we could determine groups of tropho-species having the biggest effects on each component. Finally, we did the simulations again with higher nitrogen leaching rates (from 0.1 to 0.5) to investigate how these group effects varied from high to low fertility.

\singlespacing

\begin{table}[H]
\centering
\begin{tabular}{p{1.5cm} p{2.2cm} p{1.8cm} p{5.1cm} p{2.5cm}}
\toprule
& Trait & Type & Description & Documentation \\
\hline
Observed traits & Prey capture strategy & Boolean & Web builder or not & Literature \\
 & Poison & Boolean & Use of poison to kill prey or not & Literature \\
 & Body mass & Continuous & Logarithm of the mass of an individual (in grams) & Measurement, literature* \\
 & Mobility & Categorical & 1: immobile, 2: crawling (no legs), 3: short legs, 4: long legs, 5: jumping, 6: flight & Literature  \\
 & Toughness & Categorical & 0: soft (no chitin, or few lignin), 0.5: hard, 1: has a shell (or is a seed) & Literature  \\
Latent traits & Feeding guild & Boolean & Carnivore: 1/0, detritivore: 1/0, microbivore: 1/0, herbivore: 1/0 & Literature  \\
 & Taxonomy & Continuous & Scores on the 2 PCoA axes of the taxonomy & Literature  \\ 
 & Soil vertical position & Boolean & Below soil surface : 1/0 and/or above soil surface : 1/0 & Literature \\
\hline
\end{tabular}
\caption{Description of traits and proxies of traits used in the study. \\
$\ast$ Food webs dataset and body mass measurements were provided by \citet{Digel2014}.}
\label{traits}
\end{table}

\doublespacing

All of the analyses were performed with R \citep{R} and the package "party" \citep{Tosten2006,Strobl2007, Strobl2008}. 

\section*{Results}

\begin{figure}[H]
\begin{center}
\includegraphics[width=1\textwidth]{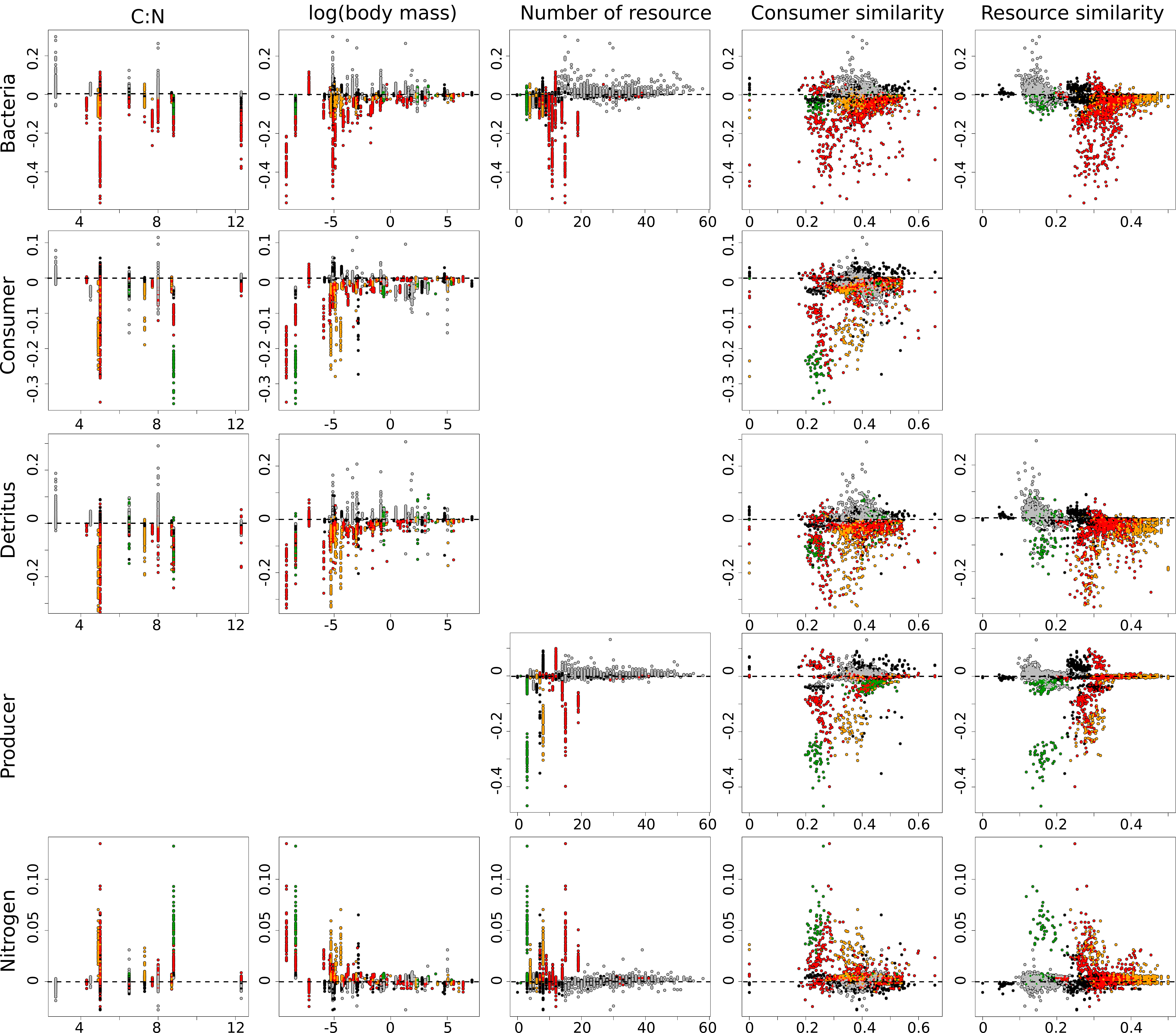}
\caption{Relationships between tropho-species properties and their \textit{per population} effect on each component. Only properties which the random forest revealed significant in the explanation of tropho-species effects are presented. Green points are herbivores, orange ones are strict detritivores, red ones are detritivores and fungivores, and grey ones are carnivores, black ones are all other species.}
\label{reg}
\end{center}
\end{figure}

Removal effects of tropho-species were most often null, but certain tropho-species had high effects in most food webs (figure \ref{reg}). Fungi biomass, however, responded greatly to the majority of tropho-species removals (75\% of effects had an absolute value > 0.1). Additionally, no removal led to the extinction of other species (species biomass > 0.0001). Tropho-species \textit{per population} and \textit{per capita} removal effects on each component were accurately predicted by their traits and trophic properties ($Rpo>0.70$), except for their effects on fungi which were highly dependent on food web identity ($Rpo=0.50$). The main properties that predict tropho-species removal effects were their number of resources, mean trophic similarity, diet, body mass, C:N ratio, mobility and toughness. 

\vspace{1.0cm}

\subsection*{Properties of tropho-species having the highest effects}
Strict herbivores and detritivorous species had, in general, negative \textit{per population} and \textit{per capita} removal effects on all components, but positive effects on nitrogen. Smallest ones (<0.08mg) with the lowest trophic similarity had the highest effects (figure \ref{reg}). Detritivores with a C:N ratio of five and detri-fungivorous species tended to have higher effects than other detritivores. They decreased the biomass of all species except some small size carnivores, and tended to increase more recalcitrant detritus than labile ones, and more fungi than bacteria. Most carnivores had positive or low removal effects on all components (except N). Small to medium strict carnivores (>0.2 mg and <1 mg) with low resource similarity and/or a high number of prey, of the order Lithobiomorpha, Sarcoptiformes and Parasitiformes (mites) had the highest positive \textit{per capita} and \textit{per population} removal effects on each component (except N) in the majority of food webs, although they decreased most of their prey presenting various properties. 

\subsection*{Change in tropho-species effects with decreasing fertility}
\begin{figure}[H]
\begin{center}
\includegraphics[width=1\textwidth]{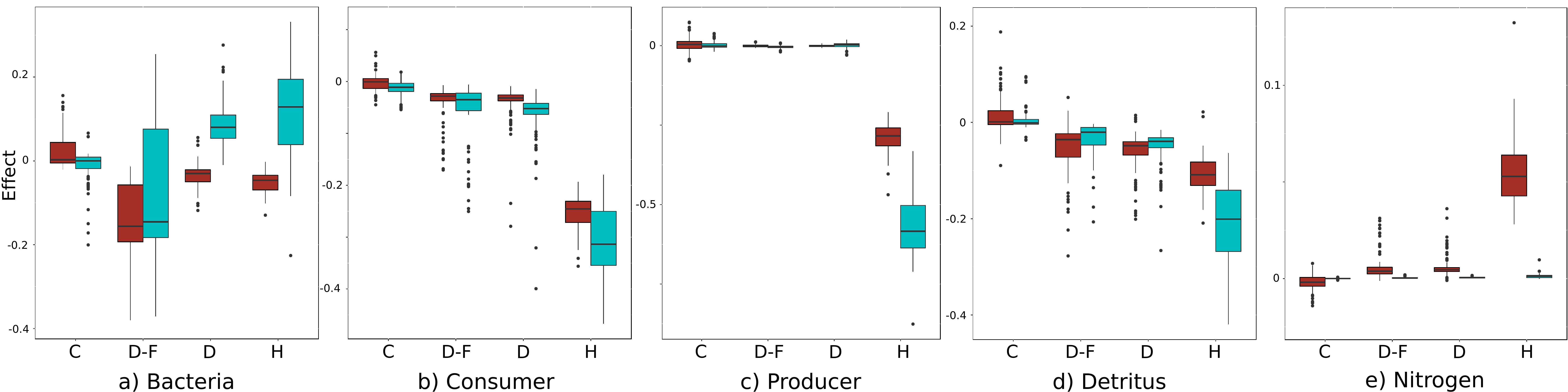}
\caption{Effects of selected groups of tropho-species on each component at high (red) and low fertility (blue). C=carnivore, D-F=detri-fungivore, D=detritivore, H=herbivore.}
\label{fertiComp}
\end{center}
\end{figure}

\begin{figure}[H]
\begin{center}
\includegraphics[width=1\textwidth]{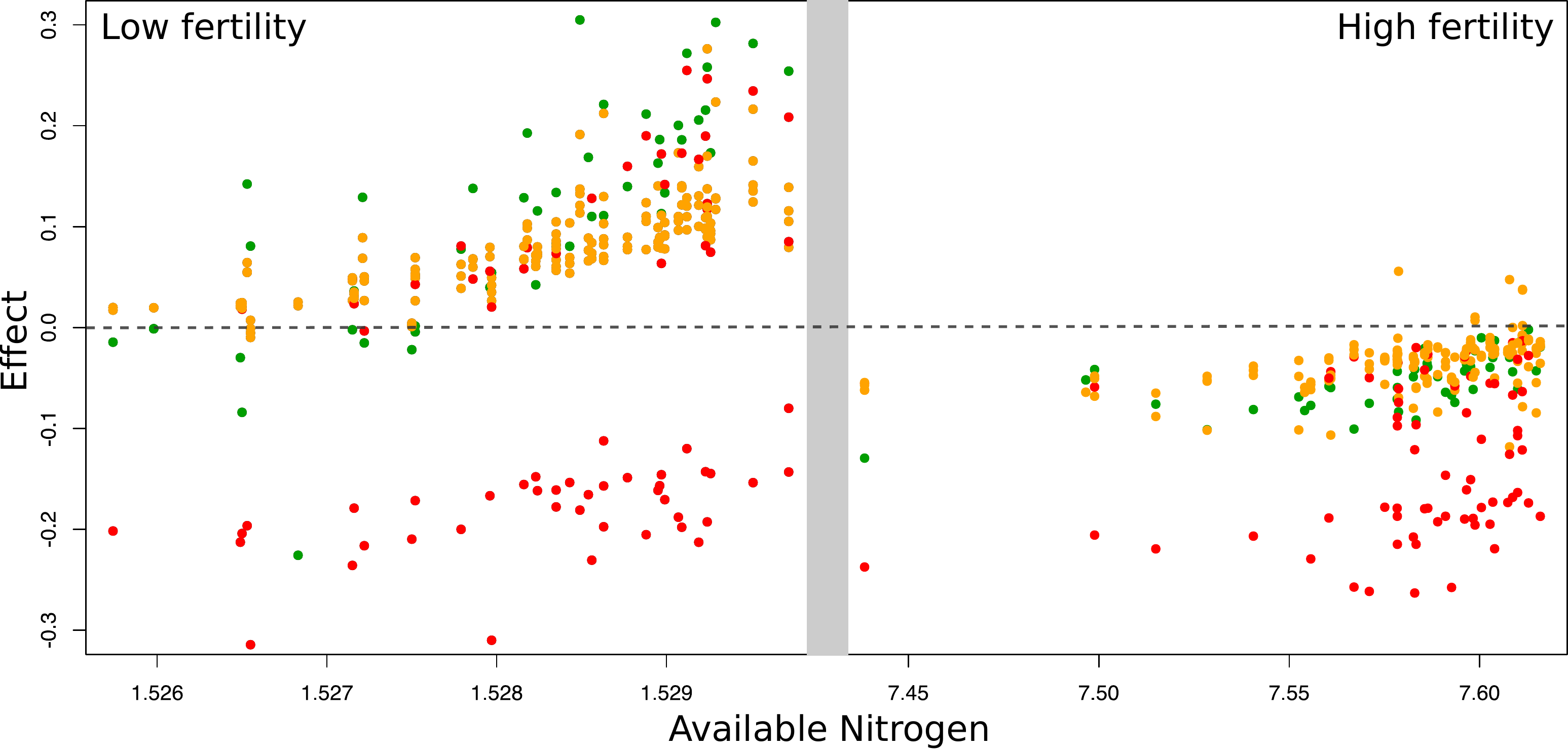}
\caption{Effects of selected groups of tropho-species on bacteria biomass according to the amount of available nitrogen in the food web at the beginning of the removal simulation. Simulations at low fertility (leaching rates of 0.5) are presented in the left part while the ones at high fertility (leaching rates of 0.1) are in the right part. Green points are herbivores, orange ones are strict detritivores, red ones are detritivores and fungivores }
\label{fertiBb}
\end{center}
\end{figure}

Extreme removal effects were higher at lower fertility in comparison to high fertility, except effects on N which were lower (figure \ref{fertiComp}). We compared the removal effects between high and low fertility of the four groups having the highest effects on each component, which were: strict herbivores, strict detritivores, detri-fungivores (<0.08 mg), strict carnivores (<1 mg, >0.2 mg). At lower fertility, strict detritivores and herbivores still had the highest removal effects on all components. However, instead of negatively affecting bacteria like at high fertility, they increased them. Carnivores, in contrast, tended to decrease bacteria instead of the opposite at high fertility. We then wondered whether there was a linear relationship between these groups effects on bacteria and food webs fertility. As we can see in the left part of figure \ref{fertiBb}, there is a positive relationship between removal effects on bacteria biomass with increasing nitrogen available. However, this relationship does not maintain when we simulated increasing fertility in the food webs (right part of figure \ref{fertiBb}). 

\vspace{1.0cm}

\section*{Discussion}

Models of community dynamics are useful tools to better understand the functioning of complex systems such as soil food webs. Using a model based on metabolic theory and stoichiometry, we revealed how species traits can influence species extinction effects. While soil food webs were particularly resistant to species extinction, we were able to outline which tropho-species affect the most ecosystem functioning. However, we found that the determination of species effects based on traits and topological position was also contingent on community properties. \\

We identified several traits that underlie species effects on other species, detritus and nitrogen within the community. As expected, species diet was the main factor explaining differences in species effects. Herbivores had the greatest negative effects on the overall food web by decreasing producers biomass and hence other herbivores, but increased nitrogen availability by limiting plant nitrogen uptake. Since producers generate the litter that supports a detritus-based food web, herbivores decreased the biomass of detritivores and their consumers. Similarly, detritivores had high negative effects on the overall food web by consuming detritus, and even more when they also consumed fungi, which agrees with previous findings \citep{Huhta1998,Laakso1999}. Body mass was the second most important trait explaining species effects. Non-carnivores with low body mass had greater effects on the various food web components than larger species with the same diet. Species with low body mass have greater metabolic rates per unit of biomass, and therefore consume a greater amount of resource per unit of biomass than larger species. Herbivorous nematodes and omnivorous enchytraeids in particular, which have the lowest body mass, had the greatest effects on the overall food web, similar to findings from \citet{Hunt2002}. However, in contrast to results from theoretical and microcosm experiments \citep{Huhta1998,Laakso1999,Setala2002}, we found negative effects of enchytraeids on producers. In our dataset, enchytraeids also consumed plants and microbes, while in the reality they may mostly consume detritus and then have greater positive effects on nitrogen mineralization and plant growth. Because body mass is also related to trophic position, carnivores with the greatest effects were not necessarily the smallest \citep{Schneider2012a, Laigle2018}. For instance, carnivorous mites had important effects because they consumed small intermediate trophic level species, subsequently increasing their resources. Species toughness and mobility were also found to be important traits underlying species effects. However, we presume that this contribution mainly stems from the fact that nematodes and enchytraeids are soft and have low mobility. In addition, the contribution of species C:N in food web dynamics was difficult to describe and may be related to species taxonomy and trophic position. We did not find a clear evidence that species with high C:N had higher effects on nitrogen mineralization, as expected. Thus, mainly the combination of body mass and diet explained observed species effects on the overall food web. \\


We considered observed interactions in real communities, allowing us to show that species effects are also contingent on network structure. Past studies that investigated species effects used simplified soil food webs \citep{Hunt1987,DeRuiter1993,Hunt2002} or based their work on food webs in which interactions were determined based on species' body mass only \citep{Brose2005, Berlow2009,Curtsdotter2011, Berg2015}. However, the network structure is not necessarily realistic in the earlier study, while results essentially derive from the community body mass distribution in the latter one. In contrast to our study, \citet{Berlow2009,Berg2015} and \citet{Wang2018} found great effects of large body mass species. In addition, high effects of small species were found positive by \citet{Curtsdotter2011} and \citet{Berg2015}, while we found them mainly negative, except on their relative consumers. These contrasting results show that body mass structure may greatly affect results \citep{Riede2011}, but also that past studies may underestimate negative indirect effects arising from more complex food webs including more diverse trophic levels. In addition, we found that trophic similarity explained a great proportion of species effects. Prey species with a low consumer similarity had significant effects on their relative consumers as they comprised a large part of their consumers diet \citep{Eklof2006, Montoya2009}. This property can also explain the important effects of nematodes and enchytraeids, which had the lowest consumer similarity among all herbivores and detritivores. Carnivores with the greatest effects consumed a wide range of resources, as also found by \citet{Sole2001,Dunne2002} and \citet{Eklof2006}, but also have a low resource similarity and small body sizes (e.g. chilopods, mites). These carnivores directly decreased their most abundant prey, indirectly favoring less abundant ones, hence increasing food web productivity through the regulation of competitive forces \citep{Paine1969,Power1996}. Because the body mass of a carnivore partly determines its prey, its value relative to body mass of other species should capture a carnivore's importance in a food web \citep{Schneider2012a}. Our results highlight the importance of connectivity for top-down effects regulation mechanisms more than bottom-up effects as demonstrated by \citet{Eklof2006} and \citet{Curtsdotter2011}. Further, \citet{Laigle2018} showed how functionally close species tend to be trophically similar. However, the low importance of functional distance in our study highlights that we are missing traits that adequately represent trophic similarity. All together, these results suggest that the determination of species effects and the identification of keystone species must refer to a specific network structure. \\

The precision and realism of our detritus-based stoichiometrically explicit model highlight the importance of indirect effects and demonstrate the variability in species effects based on soil fertility and litter composition. While the importance of indirect effects is commonly accepted, their consideration is often limited to competition regulation \citep{Brose2005,Berlow2009, Montoya2009}. We showed how the consideration of C and N cycling dynamic can be essential at understanding variability in species effects. For instance, at low fertility, herbivores have positive effects on bacteria because producers and bacteria compete for nitrogen, whereas we observe the opposite at high fertility. In addition, overall effects of species on microbes were difficult to understand because effects from their excretion are added to their consumption or competition effects. Further, these effects vary with fertility as it affects species resources' C:N which in turn affects their excretions' C:N. Changing fertility has then the potential to change a mutualistic relationship between a species and microbes to an antagonistic one. While changing effects of species extinction in various food webs have been mainly explained by food web structure \citep{Eklof2006,Petchey2008,Curtsdotter2011,Riede2011}, we add that the determination of species effects depend also on the level of soil fertility or basal resource stoichiometry in a given food web. \\


Conclusions from such numerical experiments are contingent on the model structure, with some results being more robust than others to its assumptions. Low resolution of basal species is one of the limitation that should be overpass to further improve the realism of our study. Indeed, finer resolution of microbial species and their interactions could better specify the role of microbivores and herbivores. Some herbivores could favor plants that produce a nitrogen-rich litter, which would then favor mineralization. In addition, the coarse resolution of detritus pools and interactions between detritivores hide potential complementarity among them. Complementarity and facilitation occurring throughout the decomposition process were documented by several studies and may surpass competition \citep{DeOliveira2010, Hedde2010}. Mesodetritivores, mesofungivores and microbes are the main nitrogen producers in soil food webs, and thus their interactions deserve to be more documented \citep{Brose2014}. Several parameters also lack precision because of a lack of data, such as assimilation efficiency of consumers. We also did not take into account resource handling time and prey ability to escape consumption (refuge) that may be important factors in population dynamics of soil organisms \citet{Elliott1980,Hunt1987,Pawar2015}. But even if our model lack some mechanisms, we greatly improved the realism of soil models by incorporating detritus, stoichiometry and compensatory mechanisms \citep{Buchkowski2016} leading to contrasting results from other studies. Moreover, even if we had 48 food webs from various ecosystems, we only studied temperate forest soil food webs. It would be interesting to test the model with great changes in parameters and food web structure, for instance with datasets from other biomes where body mass distribution can be highly different (e.g. giant lumbricidae in Australia). \\

In this study, we showed that species effects depend on a combination of traits relative to the traits of other species comprising the community. We also highlighted that explicitly accounting for carbon and nitrogen flows emphasizes the importance of feedbacks and indirect effects in detritus based food webs. We highlighted for the first time how important it is to assess a species contribution to ecosystem functioning in regard to the community structure and ecosystem properties. Metrics of food web topological and functional structure are great tools to improve our understanding of soil community functioning.

\singlespacing

\bibliography{library}
\bibliographystyle{plainnat}

\newgeometry{top=0.2cm,bottom=0.5cm,right=0.5cm,left=0.5cm}

\section*{Appendix 1}

\begin{table}[H]
\thispagestyle{empty}
\centering
\begin{tabular}{p{0.5cm} p{0.5cm} p{0.7cm} p{0.7cm} p{0.5cm} p{1.5cm} p{2.0cm} p{2.5cm} p{2.0cm} p{1.5cm} p{1.5cm}}
\hline
FW & RS & RT & NL & C & Generality & Vulnerability & Chain length & Omnivory & Similarity & Modularity \\
\hline
1 & 115 & 78 & 1630 & 0.12 & 17.42 & 10.93 & 1.94 & 0.27 & 0.98 & 0.18 \\
2 & 147 & 89 & 3122 & 0.14 & 24.85 & 13.19 & 2.11 & 0.35 & 0.98 & 0.21 \\
3 & 116 & 80 & 1650 & 0.12 & 15.8 & 10.16 & 1.99 & 0.32 & 0.98 & 0.2 \\
4 & 119 & 76 & 1641 & 0.12 & 16.99 & 10.22 & 2.1 & 0.33 & 0.96 & 0.22 \\
5 & 120 & 80 & 1666 & 0.12 & 19.34 & 9.31 & 1.93 & 0.3 & 0.97 & 0.2 \\
6 & 114 & 85 & 1546 & 0.12 & 16.85 & 9.09 & 1.93 & 0.32 & 0.96 & 0.25 \\
7 & 123 & 82 & 1729 & 0.11 & 18.64 & 9.37 & 1.83 & 0.26 & 0.96 & 0.2 \\
8 & 130 & 84 & 2081 & 0.12 & 18.16 & 11.29 & 2.07 & 0.35 & 0.98 & 0.23 \\
9 & 128 & 84 & 2165 & 0.13 & 20.44 & 11.83 & 2.08 & 0.37 & 0.98 & 0.2 \\
10 & 114 & 72 & 1679 & 0.13 & 18.05 & 9.87 & 2.17 & 0.31 & 0.98 & 0.22 \\
11 & 139 & 87 & 2636 & 0.14 & 22.36 & 12.73 & 2.08 & 0.39 & 0.98 & 0.23 \\
12 & 105 & 72 & 1428 & 0.13 & 16.36 & 8.56 & 2.08 & 0.33 & 0.97 & 0.25 \\
13 & 117 & 71 & 1664 & 0.12 & 16.17 & 9.51 & 2.1 & 0.37 & 0.97 & 0.32 \\
14 & 114 & 80 & 1157 & 0.09 & 16.56 & 7.38 & 1.75 & 0.24 & 0.98 & 0.27 \\
15 & 108 & 76 & 1197 & 0.1 & 14.16 & 8.97 & 1.82 & 0.27 & 0.97 & 0.21 \\
16 & 148 & 84 & 2634 & 0.12 & 21.93 & 12.01 & 1.97 & 0.31 & 0.98 & 0.2 \\
17 & 128 & 83 & 1310 & 0.08 & 17.31 & 8.72 & 1.78 & 0.25 & 0.97 & 0.22 \\
18 & 120 & 81 & 1555 & 0.11 & 16.45 & 9.5 & 1.96 & 0.3 & 0.96 & 0.28 \\
19 & 117 & 79 & 1454 & 0.11 & 12.33 & 10.01 & 2.01 & 0.31 & 0.97 & 0.23 \\
20 & 110 & 75 & 975 & 0.08 & 13.67 & 7.1 & 1.72 & 0.21 & 0.98 & 0.3 \\
21 & 162 & 86 & 2849 & 0.11 & 24.04 & 13.22 & 2.13 & 0.38 & 0.98 & 0.25 \\
22 & 136 & 87 & 2091 & 0.11 & 20.53 & 11.31 & 2.01 & 0.36 & 0.96 & 0.22 \\
23 & 85 & 66 & 718 & 0.1 & 9.54 & 6.76 & 1.89 & 0.29 & 0.97 & 0.3 \\
24 & 126 & 83 & 1692 & 0.11 & 18.91 & 10.14 & 1.95 & 0.32 & 0.97 & 0.26 \\
25 & 137 & 83 & 2299 & 0.12 & 22.02 & 11.27 & 2.02 & 0.36 & 0.98 & 0.25 \\
26 & 138 & 86 & 1880 & 0.1 & 20.45 & 10.37 & 1.84 & 0.27 & 0.97 & 0.25 \\
27 & 126 & 76 & 2065 & 0.13 & 20.23 & 10.94 & 2.07 & 0.41 & 0.97 & 0.26 \\
28 & 133 & 87 & 1894 & 0.11 & 19.24 & 10.58 & 2.09 & 0.37 & 0.97 & 0.29 \\
29 & 140 & 87 & 1740 & 0.09 & 19.05 & 9.54 & 1.94 & 0.29 & 0.97 & 0.31 \\
30 & 131 & 76 & 1918 & 0.11 & 17.38 & 10.35 & 2.05 & 0.36 & 0.98 & 0.31 \\
31 & 154 & 89 & 2964 & 0.12 & 22.29 & 12.38 & 1.95 & 0.31 & 0.97 & 0.23 \\
32 & 142 & 89 & 2839 & 0.14 & 22.82 & 14.15 & 2.15 & 0.4 & 0.98 & 0.21 \\
33 & 98 & 66 & 1109 & 0.12 & 11.85 & 9.45 & 2.23 & 0.34 & 0.98 & 0.25 \\
34 & 87 & 66 & 566 & 0.07 & 9.22 & 6.66 & 1.83 & 0.24 & 0.97 & 0.28 \\
35 & 117 & 70 & 2006 & 0.15 & 16 & 11.85 & 2.3 & 0.43 & 0.97 & 0.24 \\
36 & 114 & 76 & 1417 & 0.11 & 14.44 & 9.17 & 2.04 & 0.32 & 0.97 & 0.28 \\
37 & 95 & 67 & 1076 & 0.12 & 13.5 & 8.73 & 1.83 & 0.24 & 0.97 & 0.2 \\
38 & 121 & 76 & 1796 & 0.12 & 14.14 & 10.65 & 2.19 & 0.35 & 0.98 & 0.28 \\
39 & 126 & 78 & 2609 & 0.16 & 19.22 & 13.02 & 2.18 & 0.38 & 0.98 & 0.21 \\
40 & 112 & 72 & 1734 & 0.14 & 15.88 & 10.12 & 2.16 & 0.39 & 0.97 & 0.21 \\
41 & 83 & 64 & 921 & 0.13 & 10.97 & 8.19 & 2.04 & 0.33 & 0.97 & 0.22 \\
42 & 118 & 72 & 1776 & 0.13 & 16.82 & 12.54 & 2.26 & 0.37 & 0.98 & 0.13 \\
43 & 95 & 67 & 1026 & 0.11 & 11.47 & 8.66 & 2.03 & 0.31 & 0.97 & 0.24 \\
44 & 83 & 70 & 819 & 0.12 & 9.67 & 7.54 & 2.09 & 0.31 & 0.97 & 0.24 \\
45 & 112 & 73 & 1856 & 0.15 & 13.72 & 10.54 & 2.14 & 0.39 & 0.97 & 0.2 \\
46 & 112 & 84 & 1568 & 0.12 & 15.42 & 10.47 & 2.07 & 0.36 & 0.97 & 0.24 \\
47 & 114 & 84 & 1779 & 0.14 & 18.37 & 12.05 & 1.93 &2.07 & 0.97 & 0.22 \\
48 & 112 & 74 & 1711 & 0.14 & 15.78 & 10.37 & 2.12 & 0.35 & 0.98 & 0.23 \\
\hline
\end{tabular}
\caption{Properties of the food webs studies: FW is the food web number, RS is the number of species, RT is the number of tropho-species, NL is the number of links. See \citep{Laigle2018} for more detailed description of the properties.}
\label{FW}
\end{table}

\thispagestyle{empty}
 
\newpage

\begin{landscape}

\section*{Appendix 2}

\begin{longtable}{p{0.8cm} p{1.0cm} p{1.3cm} p{0.5cm} p{1.0cm} p{0.8cm} p{0.8cm} p{0.8cm} p{0.6cm} p{0.5cm} p{0.7cm} p{0.7cm} p{3.5cm} p{2.0cm} p{2.2cm} p{2.0cm}}
\hline
Group & Tough & Mobility & Web & Poison & Mass range & Above & Below & Carn & Det & Herb & Fung & Order & Class & Phylum & Kingdom \\
1 & 0 & 0 & 0 & 0 & 0 & 0 & 1 & 0 & 0 & 0 & 0 & Algae & Algae & Cyanobacteria & Bacteria \\
2 & 0 & 0 & 0 & 0 & 1 & 0 & 1 & 0 & 1 & 0 & 0 & Bacteria & Bacteria & Bacteria & Bacteria \\
3 & 0 & 0 & 0 & 0 & 0 & 1 & 1 & 0 & 0 & 0 & 0 & Detritus & Detritus & Detritus & Detritus \\
3 & 0 & 0 & 0 & 0 & 0 & 1 & 0 & 0 & 0 & 0 & 0 & Detritus & Detritus & Detritus & Detritus \\
4 & 0 & 0 & 0 & 0 & 0 & 0 & 1 & 0 & 1 & 0 & 0 & Fungi & Fungi & Fungi & Fungi \\
6 & 0 & 0 & 0 & 0 & 0 & 0 & 1 & 0 & 0 & 0 & 0 & Plantae & Plantae & Plantae & Plantae \\
6 & 0 & 0 & 0 & 0 & 0 & 1 & 0 & 0 & 0 & 0 & 0 & Plantae & Plantae & Plantae & Plantae \\
7 & 0 & 0 & 0 & 0 & 5 & 0 & 1 & 0 & 0 & 0 & 1 & Protozoa & Protozoa & Protozoa & Protozoa \\
8 & 0 & 1 & 0 & 0 & 10 & 0 & 1 & 1 & 1 & 0 & 1 & Haplotaxida & Clitellata & Annelida & Animalia \\
9 & 0 & 1 & 0 & 0 & 6 & 1 & 1 & 1 & 1 & 1 & 1 & Haplotaxida & Clitellata & Annelida & Animalia \\
10 & 0 & 1 & 0 & 0 & 6 & 0 & 1 & 1 & 0 & 0 & 0 & Nematoda & Nematoda & Nematoda & Animalia \\
11 & 0 & 1 & 0 & 0 & 6 & 0 & 1 & 0 & 0 & 1 & 0 & Nematoda & Nematoda & Nematoda & Animalia \\
12 & 0 & 1 & 0 & 0 & 6 & 0 & 1 & 0 & 0 & 0 & 1 & Nematoda & Nematoda & Nematoda & Animalia \\
13 & 0 & 1 & 0 & 0 & 6 & 0 & 1 & 1 & 1 & 1 & 1 & Nematoda & Nematoda & Nematoda & Animalia \\
14 & 0 & 1 & 0 & 0 & 10 & 1 & 1 & 0 & 0 & 1 & 0 & Pulmonata & Gastropoda & Mollusca & Animalia \\
15 & 0 & 1 & 0 & 0 & 10 & 1 & 1 & 0 & 1 & 0 & 0 & Pulmonata & Gastropoda & Mollusca & Animalia \\
16 & 0 & 1 & 0 & 0 & 9 & 1 & 1 & 0 & 1 & 0 & 0 & Pulmonata & Gastropoda & Mollusca & Animalia \\
17 & 0 & 1 & 0 & 0 & 10 & 1 & 1 & 1 & 1 & 0 & 0 & Pulmonata & Gastropoda & Mollusca & Animalia \\
18 & 0 & 1 & 0 & 0 & 9 & 1 & 1 & 1 & 1 & 0 & 0 & Pulmonata & Gastropoda & Mollusca & Animalia \\
19 & 0 & 1 & 0 & 0 & 10 & 1 & 1 & 0 & 1 & 1 & 0 & Pulmonata & Gastropoda & Mollusca & Animalia \\
20 & 0 & 1 & 0 & 0 & 10 & 1 & 1 & 0 & 0 & 1 & 0 & Pulmonata & Gastropoda & Mollusca & Animalia \\
21 & 0 & 1 & 0 & 0 & 10 & 1 & 1 & 1 & 0 & 1 & 1 & Pulmonata & Gastropoda & Mollusca & Animalia \\
22 & 0 & 1 & 0 & 0 & 10 & 1 & 1 & 0 & 1 & 0 & 1 & Pulmonata & Gastropoda & Mollusca & Animalia \\
23 & 0 & 1 & 0 & 0 & 10 & 1 & 1 & 0 & 1 & 1 & 1 & Pulmonata & Gastropoda & Mollusca & Animalia \\
24 & 0 & 1 & 0 & 0 & 9 & 1 & 1 & 0 & 1 & 1 & 1 & Pulmonata & Gastropoda & Mollusca & Animalia \\
25 & 0 & 2 & 0 & 0 & 7 & 0 & 1 & 0 & 1 & 0 & 0 & Entomobryomorpha & Collembola & Arthropoda & Animalia \\
26 & 0 & 2 & 0 & 0 & 8 & 1 & 0 & 0 & 1 & 0 & 0 & Entomobryomorpha & Collembola & Arthropoda & Animalia \\
27 & 0 & 2 & 0 & 0 & 7 & 0 & 1 & 1 & 0 & 0 & 1 & Entomobryomorpha & Collembola & Arthropoda & Animalia \\
28 & 0 & 2 & 0 & 0 & 8 & 0 & 1 & 1 & 0 & 0 & 1 & Entomobryomorpha & Collembola & Arthropoda & Animalia \\
29 & 0 & 2 & 0 & 0 & 7 & 0 & 1 & 1 & 0 & 0 & 0 & Poduromorpha & Collembola & Arthropoda & Animalia \\
29 & 0 & 2 & 0 & 0 & 7 & 1 & 1 & 1 & 0 & 0 & 0 & Poduromorpha & Collembola & Arthropoda & Animalia \\
30 & 0 & 2 & 0 & 0 & 8 & 1 & 1 & 1 & 0 & 0 & 0 & Poduromorpha & Collembola & Arthropoda & Animalia \\
30 & 0 & 2 & 0 & 0 & 8 & 0 & 1 & 1 & 0 & 0 & 0 & Poduromorpha & Collembola & Arthropoda & Animalia \\
31 & 0 & 2 & 0 & 0 & 7 & 0 & 1 & 1 & 0 & 0 & 1 & Poduromorpha & Collembola & Arthropoda & Animalia \\
31 & 0 & 2 & 0 & 0 & 7 & 1 & 0 & 1 & 0 & 0 & 1 & Poduromorpha & Collembola & Arthropoda & Animalia \\
31 & 0 & 2 & 0 & 0 & 7 & 1 & 1 & 1 & 0 & 0 & 1 & Poduromorpha & Collembola & Arthropoda & Animalia \\
32 & 0 & 2 & 0 & 0 & 8 & 1 & 0 & 1 & 0 & 0 & 1 & Poduromorpha & Collembola & Arthropoda & Animalia \\
32 & 0 & 2 & 0 & 0 & 8 & 0 & 1 & 1 & 0 & 0 & 1 & Poduromorpha & Collembola & Arthropoda & Animalia \\
33 & 0 & 2 & 0 & 0 & 7 & 0 & 1 & 0 & 1 & 0 & 1 & Poduromorpha & Collembola & Arthropoda & Animalia \\
34 & 0 & 2 & 0 & 0 & 7 & 0 & 1 & 1 & 1 & 0 & 1 & Poduromorpha & Collembola & Arthropoda & Animalia \\
35 & 0 & 2 & 0 & 0 & 9 & 0 & 1 & 1 & 1 & 1 & 1 & Rhabdura & Diplura & Arthropoda & Animalia \\
36 & 0 & 2 & 0 & 0 & 10 & 0 & 1 & 1 & 1 & 1 & 1 & Rhabdura & Diplura & Arthropoda & Animalia \\
37 & 0 & 2 & 0 & 0 & 9 & 0 & 1 & 0 & 1 & 1 & 1 & Scolopendromorpha & Chilopoda & Arthropoda & Animalia \\
38 & 0 & 2 & 0 & 0 & 8 & 0 & 1 & 0 & 1 & 1 & 1 & Scolopendromorpha & Chilopoda & Arthropoda & Animalia \\
39 & 0 & 2 & 0 & 0 & 8 & 0 & 1 & 0 & 1 & 1 & 1 & Scutigeromorpha & Symphyla & Arthropoda & Animalia \\
40 & 0 & 2 & 0 & 0 & 7 & 0 & 1 & 0 & 1 & 1 & 1 & Scutigeromorpha & Symphyla & Arthropoda & Animalia \\
41 & 0 & 2 & 0 & 0 & 9 & 0 & 1 & 0 & 1 & 1 & 1 & Scutigeromorpha & Symphyla & Arthropoda & Animalia \\
42 & 0 & 3 & 0 & 0 & 7 & 1 & 0 & 0 & 1 & 1 & 0 & Symphypleona & Collembola & Arthropoda & Animalia \\
43 & 0 & 4 & 0 & 0 & 7 & 1 & 0 & 0 & 1 & 0 & 0 & Entomobryomorpha & Collembola & Arthropoda & Animalia \\
43 & 0 & 4 & 0 & 0 & 7 & 1 & 1 & 0 & 1 & 0 & 0 & Entomobryomorpha & Collembola & Arthropoda & Animalia \\
43 & 0 & 4 & 0 & 0 & 7 & 0 & 1 & 0 & 1 & 0 & 0 & Entomobryomorpha & Collembola & Arthropoda & Animalia \\
44 & 0 & 4 & 0 & 0 & 9 & 1 & 1 & 0 & 1 & 0 & 0 & Entomobryomorpha & Collembola & Arthropoda & Animalia \\
45 & 0 & 4 & 0 & 0 & 8 & 1 & 1 & 0 & 1 & 0 & 0 & Entomobryomorpha & Collembola & Arthropoda & Animalia \\
45 & 0 & 4 & 0 & 0 & 8 & 0 & 1 & 0 & 1 & 0 & 0 & Entomobryomorpha & Collembola & Arthropoda & Animalia \\
45 & 0 & 4 & 0 & 0 & 8 & 1 & 0 & 0 & 1 & 0 & 0 & Entomobryomorpha & Collembola & Arthropoda & Animalia \\
46 & 0 & 4 & 0 & 0 & 7 & 1 & 1 & 1 & 1 & 0 & 0 & Entomobryomorpha & Collembola & Arthropoda & Animalia \\
47 & 0 & 4 & 0 & 0 & 7 & 1 & 0 & 1 & 0 & 0 & 1 & Entomobryomorpha & Collembola & Arthropoda & Animalia \\
47 & 0 & 4 & 0 & 0 & 7 & 0 & 1 & 1 & 0 & 0 & 1 & Entomobryomorpha & Collembola & Arthropoda & Animalia \\
47 & 0 & 4 & 0 & 0 & 7 & 1 & 1 & 1 & 0 & 0 & 1 & Entomobryomorpha & Collembola & Arthropoda & Animalia \\
48 & 0 & 4 & 0 & 0 & 8 & 1 & 1 & 1 & 1 & 0 & 1 & Entomobryomorpha & Collembola & Arthropoda & Animalia \\
49 & 0 & 4 & 0 & 0 & 7 & 1 & 1 & 0 & 1 & 1 & 0 & Entomobryomorpha & Collembola & Arthropoda & Animalia \\
50 & 0 & 4 & 0 & 0 & 8 & 1 & 1 & 0 & 1 & 0 & 0 & Neelipleona & Collembola & Arthropoda & Animalia \\
51 & 0 & 4 & 0 & 0 & 8 & 1 & 0 & 1 & 1 & 0 & 1 & Neelipleona & Collembola & Arthropoda & Animalia \\
52 & 0 & 4 & 0 & 0 & 7 & 1 & 0 & 1 & 0 & 0 & 0 & Poduromorpha & Collembola & Arthropoda & Animalia \\
53 & 0 & 4 & 0 & 0 & 8 & 1 & 0 & 0 & 1 & 0 & 0 & Symphypleona & Collembola & Arthropoda & Animalia \\
54 & 0 & 4 & 0 & 0 & 8 & 1 & 0 & 0 & 1 & 1 & 0 & Symphypleona & Collembola & Arthropoda & Animalia \\
54 & 0 & 4 & 0 & 0 & 8 & 1 & 1 & 0 & 1 & 1 & 0 & Symphypleona & Collembola & Arthropoda & Animalia \\
55 & 0 & 4 & 0 & 0 & 7 & 1 & 0 & 0 & 1 & 1 & 0 & Symphypleona & Collembola & Arthropoda & Animalia \\
55 & 0 & 4 & 0 & 0 & 7 & 1 & 1 & 0 & 1 & 1 & 0 & Symphypleona & Collembola & Arthropoda & Animalia \\
55 & 0 & 4 & 0 & 0 & 7 & 0 & 1 & 0 & 1 & 1 & 0 & Symphypleona & Collembola & Arthropoda & Animalia \\
56 & 0 & 4 & 0 & 0 & 9 & 1 & 0 & 0 & 1 & 1 & 0 & Symphypleona & Collembola & Arthropoda & Animalia \\
57 & 0.5 & 2 & 0 & 1 & 10 & 1 & 0 & 1 & 0 & 0 & 0 & Geophilomorpha & Chilopoda & Arthropoda & Animalia \\
58 & 0.5 & 2 & 0 & 0 & 8 & 1 & 1 & 0 & 1 & 0 & 1 & Isopoda & Malacostraca & Arthropoda & Animalia \\
59 & 0.5 & 2 & 0 & 0 & 7 & 0 & 1 & 1 & 0 & 0 & 0 & Parasitiformes & Arachnida & Arthropoda & Animalia \\
60 & 0.5 & 2 & 0 & 0 & 8 & 0 & 1 & 1 & 0 & 0 & 0 & Parasitiformes & Arachnida & Arthropoda & Animalia \\
61 & 0.5 & 2 & 0 & 0 & 8 & 0 & 1 & 0 & 0 & 0 & 1 & Sarcoptiformes & Arachnida & Arthropoda & Animalia \\
62 & 0.5 & 2 & 0 & 0 & 8 & 0 & 1 & 0 & 1 & 1 & 1 & Sarcoptiformes & Arachnida & Arthropoda & Animalia \\
63 & 0.5 & 3 & 0 & 1 & 10 & 1 & 0 & 1 & 0 & 0 & 0 & Araneae & Arachnida & Arthropoda & Animalia \\
64 & 0.5 & 3 & 1 & 1 & 10 & 1 & 0 & 1 & 0 & 0 & 0 & Araneae & Arachnida & Arthropoda & Animalia \\
65 & 0.5 & 3 & 1 & 1 & 9 & 1 & 0 & 1 & 0 & 0 & 0 & Araneae & Arachnida & Arthropoda & Animalia \\
66 & 0.5 & 3 & 1 & 1 & 8 & 1 & 0 & 1 & 0 & 0 & 0 & Araneae & Arachnida & Arthropoda & Animalia \\
67 & 0.5 & 3 & 0 & 1 & 10 & 1 & 0 & 1 & 0 & 0 & 0 & Chilopoda & Chilopoda & Arthropoda & Animalia \\
68 & 0.5 & 3 & 0 & 0 & 9 & 1 & 0 & 0 & 1 & 0 & 1 & Chordeumatida & Diplopoda & Arthropoda & Animalia \\
69 & 0.5 & 3 & 0 & 0 & 10 & 1 & 0 & 0 & 1 & 0 & 1 & Chordeumatida & Diplopoda & Arthropoda & Animalia \\
69 & 0.5 & 3 & 0 & 0 & 10 & 1 & 1 & 0 & 1 & 0 & 1 & Chordeumatida & Diplopoda & Arthropoda & Animalia \\
70 & 0.5 & 3 & 0 & 0 & 10 & 1 & 0 & 1 & 0 & 0 & 0 & Coleoptera & Insecta & Arthropoda & Animalia \\
71 & 0.5 & 3 & 0 & 0 & 9 & 1 & 0 & 1 & 0 & 0 & 0 & Coleoptera & Insecta & Arthropoda & Animalia \\
72 & 0.5 & 3 & 0 & 0 & 8 & 1 & 0 & 0 & 0 & 0 & 1 & Coleoptera & Insecta & Arthropoda & Animalia \\
73 & 0.5 & 3 & 0 & 0 & 9 & 1 & 0 & 1 & 0 & 0 & 1 & Coleoptera & Insecta & Arthropoda & Animalia \\
74 & 0.5 & 3 & 0 & 1 & 10 & 1 & 0 & 1 & 0 & 0 & 0 & Geophilomorpha & Chilopoda & Arthropoda & Animalia \\
75 & 0.5 & 3 & 0 & 1 & 9 & 1 & 0 & 1 & 0 & 0 & 0 & Geophilomorpha & Chilopoda & Arthropoda & Animalia \\
76 & 0.5 & 3 & 0 & 0 & 9 & 1 & 1 & 0 & 1 & 0 & 1 & Isopoda & Malacostraca & Arthropoda & Animalia \\
77 & 0.5 & 3 & 0 & 0 & 10 & 1 & 1 & 0 & 1 & 0 & 1 & Isopoda & Malacostraca & Arthropoda & Animalia \\
78 & 0.5 & 3 & 0 & 0 & 8 & 1 & 1 & 0 & 1 & 0 & 1 & Isopoda & Malacostraca & Arthropoda & Animalia \\
79 & 0.5 & 3 & 0 & 0 & 10 & 1 & 1 & 1 & 1 & 0 & 1 & Isopoda & Malacostraca & Arthropoda & Animalia \\
80 & 0.5 & 3 & 0 & 0 & 10 & 1 & 0 & 0 & 1 & 0 & 1 & Julida & Diplopoda & Arthropoda & Animalia \\
81 & 0.5 & 3 & 0 & 0 & 9 & 1 & 0 & 0 & 1 & 0 & 1 & Julida & Diplopoda & Arthropoda & Animalia \\
82 & 0.5 & 3 & 0 & 0 & 9 & 1 & 0 & 1 & 1 & 0 & 1 & Julida & Diplopoda & Arthropoda & Animalia \\
83 & 0.5 & 3 & 0 & 1 & 10 & 1 & 0 & 1 & 0 & 0 & 0 & Lithobiomorpha & Chilopoda & Arthropoda & Animalia \\
84 & 0.5 & 3 & 0 & 1 & 9 & 1 & 0 & 1 & 0 & 0 & 0 & Lithobiomorpha & Chilopoda & Arthropoda & Animalia \\
85 & 0.5 & 3 & 0 & 1 & 7 & 1 & 0 & 1 & 0 & 0 & 0 & Lithobiomorpha & Chilopoda & Arthropoda & Animalia \\
86 & 0.5 & 3 & 0 & 1 & 8 & 1 & 0 & 1 & 0 & 0 & 0 & Lithobiomorpha & Chilopoda & Arthropoda & Animalia \\
87 & 0.5 & 3 & 1 & 0 & 10 & 1 & 0 & 1 & 0 & 0 & 0 & Opiliones & Arachnida & Arthropoda & Animalia \\
88 & 0.5 & 3 & 1 & 0 & 9 & 1 & 0 & 1 & 0 & 0 & 0 & Opiliones & Arachnida & Arthropoda & Animalia \\
89 & 0.5 & 3 & 0 & 0 & 8 & 0 & 1 & 1 & 0 & 0 & 0 & Parasitiformes & Arachnida & Arthropoda & Animalia \\
90 & 0.5 & 3 & 0 & 0 & 9 & 0 & 1 & 1 & 0 & 0 & 0 & Parasitiformes & Arachnida & Arthropoda & Animalia \\
91 & 0.5 & 3 & 0 & 0 & 7 & 0 & 1 & 1 & 0 & 0 & 0 & Parasitiformes & Arachnida & Arthropoda & Animalia \\
92 & 0.5 & 3 & 0 & 0 & 9 & 1 & 0 & 0 & 1 & 0 & 1 & Polydesmida & Diplopoda & Arthropoda & Animalia \\
93 & 0.5 & 3 & 0 & 0 & 10 & 1 & 0 & 0 & 1 & 0 & 1 & Polydesmida & Diplopoda & Arthropoda & Animalia \\
94 & 0.5 & 3 & 0 & 1 & 10 & 1 & 1 & 1 & 0 & 0 & 0 & Pseudoscorpionida & Arachnida & Arthropoda & Animalia \\
95 & 0.5 & 3 & 0 & 1 & 10 & 1 & 0 & 1 & 0 & 0 & 0 & Scolopendromorpha & Chilopoda & Arthropoda & Animalia \\
96 & 0.5 & 3 & 0 & 0 & 9 & 0 & 1 & 1 & 0 & 0 & 0 & Trombidiformes & Arachnida & Arthropoda & Animalia \\
97 & 0.5 & 4 & 0 & 1 & 10 & 1 & 0 & 1 & 0 & 0 & 0 & Araneae & Arachnida & Arthropoda & Animalia \\
98 & 0.5 & 4 & 0 & 1 & 9 & 1 & 0 & 1 & 0 & 0 & 0 & Araneae & Arachnida & Arthropoda & Animalia \\
99 & 0.5 & 5 & 0 & 0 & 10 & 1 & 0 & 1 & 0 & 1 & 0 & Coleoptera & Insecta & Arthropoda & Animalia \\
100 & 1 & 2 & 0 & 0 & 9 & 0 & 1 & 0 & 1 & 0 & 0 & Sarcoptiformes & Arachnida & Arthropoda & Animalia \\
101 & 1 & 1 & 0 & 0 & 10 & 1 & 0 & 0 & 0 & 1 & 0 & Coleoptera & Insecta & Arthropoda & Animalia \\
102 & 1 & 2 & 0 & 0 & 10 & 1 & 0 & 1 & 0 & 0 & 0 & Coleoptera & Insecta & Arthropoda & Animalia \\
103 & 1 & 2 & 0 & 0 & 10 & 1 & 0 & 0 & 0 & 1 & 0 & Coleoptera & Insecta & Arthropoda & Animalia \\
104 & 1 & 2 & 0 & 0 & 10 & 1 & 0 & 0 & 1 & 0 & 0 & Coleoptera & Insecta & Arthropoda & Animalia \\
105 & 1 & 2 & 0 & 0 & 10 & 1 & 0 & 1 & 1 & 0 & 0 & Coleoptera & Insecta & Arthropoda & Animalia \\
106 & 1 & 2 & 0 & 0 & 10 & 1 & 0 & 0 & 0 & 0 & 1 & Coleoptera & Insecta & Arthropoda & Animalia \\
107 & 1 & 2 & 0 & 0 & 9 & 1 & 0 & 1 & 0 & 0 & 1 & Coleoptera & Insecta & Arthropoda & Animalia \\
108 & 1 & 2 & 0 & 0 & 10 & 1 & 0 & 0 & 1 & 0 & 0 & Glomerida & Diplopoda & Arthropoda & Animalia \\
109 & 1 & 2 & 0 & 0 & 8 & 0 & 1 & 1 & 0 & 0 & 0 & Parasitiformes & Arachnida & Arthropoda & Animalia \\
110 & 1 & 2 & 0 & 0 & 8 & 0 & 1 & 0 & 1 & 0 & 0 & Sarcoptiformes & Arachnida & Arthropoda & Animalia \\
111 & 1 & 2 & 0 & 0 & 9 & 0 & 1 & 0 & 1 & 0 & 0 & Sarcoptiformes & Arachnida & Arthropoda & Animalia \\
112 & 1 & 2 & 0 & 0 & 7 & 0 & 1 & 0 & 1 & 0 & 0 & Sarcoptiformes & Arachnida & Arthropoda & Animalia \\
113 & 1 & 2 & 0 & 0 & 8 & 0 & 1 & 0 & 0 & 1 & 1 & Sarcoptiformes & Arachnida & Arthropoda & Animalia \\
114 & 1 & 2 & 0 & 0 & 7 & 0 & 1 & 0 & 0 & 0 & 1 & Sarcoptiformes & Arachnida & Arthropoda & Animalia \\
115 & 1 & 2 & 0 & 0 & 9 & 0 & 1 & 0 & 0 & 0 & 1 & Sarcoptiformes & Arachnida & Arthropoda & Animalia \\
116 & 1 & 2 & 0 & 0 & 8 & 0 & 1 & 1 & 0 & 1 & 1 & Sarcoptiformes & Arachnida & Arthropoda & Animalia \\
117 & 1 & 2 & 0 & 0 & 7 & 0 & 1 & 1 & 0 & 0 & 1 & Sarcoptiformes & Arachnida & Arthropoda & Animalia \\
118 & 1 & 2 & 0 & 0 & 9 & 0 & 1 & 1 & 0 & 0 & 1 & Sarcoptiformes & Arachnida & Arthropoda & Animalia \\
119 & 1 & 2 & 0 & 0 & 8 & 0 & 1 & 0 & 0 & 1 & 1 & Sarcoptiformes & Arachnida & Arthropoda & Animalia \\
120 & 1 & 2 & 0 & 0 & 9 & 0 & 1 & 0 & 1 & 1 & 1 & Sarcoptiformes & Arachnida & Arthropoda & Animalia \\
121 & 1 & 2 & 0 & 0 & 8 & 0 & 1 & 0 & 1 & 1 & 1 & Sarcoptiformes & Arachnida & Arthropoda & Animalia \\
122 & 1 & 2 & 0 & 0 & 7 & 0 & 1 & 0 & 1 & 1 & 1 & Sarcoptiformes & Arachnida & Arthropoda & Animalia \\
123 & 1 & 2 & 0 & 0 & 7 & 0 & 1 & 1 & 1 & 0 & 1 & Sarcoptiformes & Arachnida & Arthropoda & Animalia \\
124 & 1 & 2 & 0 & 0 & 8 & 0 & 1 & 1 & 1 & 0 & 1 & Sarcoptiformes & Arachnida & Arthropoda & Animalia \\
125 & 1 & 2 & 0 & 0 & 6 & 0 & 1 & 1 & 1 & 0 & 1 & Sarcoptiformes & Arachnida & Arthropoda & Animalia \\
126 & 1 & 2 & 0 & 0 & 8 & 0 & 1 & 1 & 1 & 1 & 1 & Sarcoptiformes & Arachnida & Arthropoda & Animalia \\
127 & 1 & 3 & 0 & 0 & 10 & 1 & 0 & 0 & 0 & 0 & 0 & Coleoptera & Insecta & Arthropoda & Animalia \\
128 & 1 & 3 & 0 & 0 & 10 & 1 & 0 & 1 & 0 & 0 & 0 & Coleoptera & Insecta & Arthropoda & Animalia \\
129 & 1 & 3 & 0 & 0 & 9 & 1 & 0 & 1 & 0 & 0 & 0 & Coleoptera & Insecta & Arthropoda & Animalia \\
130 & 1 & 3 & 0 & 0 & 10 & 1 & 0 & 0 & 0 & 1 & 0 & Coleoptera & Insecta & Arthropoda & Animalia \\
131 & 1 & 3 & 0 & 0 & 9 & 1 & 0 & 0 & 0 & 1 & 0 & Coleoptera & Insecta & Arthropoda & Animalia \\
132 & 1 & 3 & 0 & 0 & 10 & 1 & 0 & 1 & 0 & 1 & 0 & Coleoptera & Insecta & Arthropoda & Animalia \\
133 & 1 & 3 & 0 & 0 & 9 & 1 & 0 & 0 & 0 & 0 & 1 & Coleoptera & Insecta & Arthropoda & Animalia \\
134 & 1 & 3 & 0 & 0 & 9 & 1 & 0 & 1 & 0 & 0 & 1 & Coleoptera & Insecta & Arthropoda & Animalia \\
135 & 1 & 3 & 0 & 0 & 10 & 1 & 0 & 1 & 0 & 0 & 1 & Coleoptera & Insecta & Arthropoda & Animalia \\
136 & 1 & 3 & 0 & 0 & 10 & 1 & 0 & 0 & 1 & 0 & 1 & Coleoptera & Insecta & Arthropoda & Animalia \\
137 & 1 & 3 & 0 & 0 & 10 & 1 & 0 & 1 & 1 & 0 & 1 & Coleoptera & Insecta & Arthropoda & Animalia \\
138 & 1 & 3 & 0 & 0 & 10 & 1 & 0 & 0 & 1 & 1 & 1 & Coleoptera & Insecta & Arthropoda & Animalia \\
139 & 1 & 3 & 0 & 0 & 9 & 1 & 0 & 0 & 1 & 1 & 1 & Coleoptera & Insecta & Arthropoda & Animalia \\
140 & 1 & 3 & 0 & 0 & 10 & 1 & 0 & 0 & 1 & 0 & 0 & Glomerida & Diplopoda & Arthropoda & Animalia \\
141 & 1 & 3 & 0 & 0 & 9 & 1 & 0 & 0 & 1 & 0 & 0 & Glomerida & Diplopoda & Arthropoda & Animalia \\
142 & 1 & 3 & 0 & 0 & 10 & 1 & 1 & 0 & 1 & 0 & 1 & Isopoda & Malacostraca & Arthropoda & Animalia \\
143 & 1 & 3 & 0 & 0 & 9 & 1 & 1 & 0 & 1 & 0 & 1 & Isopoda & Malacostraca & Arthropoda & Animalia \\
144 & 1 & 3 & 0 & 0 & 10 & 1 & 0 & 0 & 1 & 0 & 1 & Julida & Diplopoda & Arthropoda & Animalia \\
145 & 1 & 3 & 0 & 0 & 9 & 1 & 0 & 0 & 1 & 0 & 1 & Julida & Diplopoda & Arthropoda & Animalia \\
146 & 1 & 3 & 0 & 0 & 10 & 1 & 0 & 1 & 1 & 0 & 1 & Julida & Diplopoda & Arthropoda & Animalia \\
147 & 1 & 3 & 0 & 0 & 10 & 1 & 0 & 0 & 1 & 0 & 1 & Polydesmida & Diplopoda & Arthropoda & Animalia \\
148 & 1 & 3 & 0 & 0 & 10 & 1 & 0 & 1 & 1 & 0 & 1 & Polydesmida & Diplopoda & Arthropoda & Animalia \\
149 & 1 & 3 & 0 & 0 & 8 & 1 & 0 & 0 & 1 & 0 & 1 & Polyxenida & Diplopoda & Arthropoda & Animalia \\
150 & 1 & 3 & 0 & 0 & 8 & 0 & 1 & 1 & 0 & 0 & 1 & Sarcoptiformes & Arachnida & Arthropoda & Animalia \\
151 & 1 & 3 & 0 & 0 & 7 & 0 & 1 & 1 & 0 & 0 & 1 & Sarcoptiformes & Arachnida & Arthropoda & Animalia \\
152 & 1 & 3 & 0 & 0 & 8 & 0 & 1 & 0 & 1 & 0 & 1 & Sarcoptiformes & Arachnida & Arthropoda & Animalia \\
153 & 1 & 3 & 0 & 0 & 9 & 0 & 1 & 0 & 1 & 0 & 1 & Sarcoptiformes & Arachnida & Arthropoda & Animalia \\
154 & 1 & 3 & 0 & 0 & 7 & 0 & 1 & 0 & 1 & 0 & 1 & Sarcoptiformes & Arachnida & Arthropoda & Animalia \\
155 & 1 & 3 & 0 & 0 & 9 & 0 & 1 & 1 & 1 & 0 & 1 & Sarcoptiformes & Arachnida & Arthropoda & Animalia \\
156 & 1 & 5 & 0 & 0 & 10 & 1 & 0 & 0 & 0 & 0 & 0 & Coleoptera & Insecta & Arthropoda & Animalia \\
157 & 1 & 5 & 0 & 0 & 9 & 1 & 0 & 1 & 0 & 0 & 0 & Coleoptera & Insecta & Arthropoda & Animalia \\
158 & 1 & 5 & 0 & 0 & 10 & 1 & 0 & 0 & 0 & 1 & 0 & Coleoptera & Insecta & Arthropoda & Animalia \\
159 & 1 & 5 & 0 & 0 & 9 & 1 & 0 & 0 & 0 & 1 & 0 & Coleoptera & Insecta & Arthropoda & Animalia \\
160 & 1 & 5 & 0 & 0 & 9 & 1 & 0 & 1 & 0 & 1 & 0 & Coleoptera & Insecta & Arthropoda & Animalia \\
161 & 1 & 5 & 0 & 0 & 10 & 1 & 0 & 1 & 0 & 1 & 0 & Coleoptera & Insecta & Arthropoda & Animalia \\
\hline
\caption{Trait values and taxonomy of the tropho-species composing the 48 food webs.}
\label{TS}
\end{longtable}

\thispagestyle{empty}

\end{landscape}

\end{document}